\documentclass{SCIS2025}

\begin{document}
\ArticleType{LETTER}
\Year{2024}
\Month{}
\Vol{}
\No{}
\DOI{}
\ArtNo{}
\ReceiveDate{}
\ReviseDate{}
\AcceptDate{}
\OnlineDate{}
\AuthorMark{}
\AuthorCitation{}


\title{AudioCIL: A Python Toolbox for Audio Class-Incremental Learning with Multiple Scenes}{Qisheng Xu, Yulin Sun, Yi Su, et al. AudioCIL: A Python Toolbox for Audio Class-Incremental Learning with Multiple Scenes}

	\author[1,2]{Qisheng Xu}{}
    \author[1,2]{Yulin Sun}{}
    \author[1,2]{Yi Su}{}
    \author[1,2]{Qian Zhu}{}
    \author[1,2]{Xiaoyi Tan}{}
    \author[1,2]{\\Hongyu Wen}{}
    \author[1,2]{Zijian Gao}{}
    \author[1,2]{Kele Xu}{{xukelele@163.com}}
    \author[1,2]{Yong Dou}{}
    \author[1,2]{Dawei Feng}{}


\address[1]{National University of Defense Technology, Changsha {\rm 410073}, China}
\address[2]{College of Computer Science and Technology, Changsha {\rm 410073}, China}

\maketitle

\begin{multicols}{2}

Deep learning, with its robust automatic feature extraction capabilities, has demonstrated significant success in audio signal processing. Typically, these methods rely on static, pre-collected large-scale datasets for training, performing well on a fixed number of classes.
However, the ever-changing real world, with new audio classes emerging from streaming or temporary availability due to privacy, requires models that can learn incrementally without losing previous knowledge, such as a speech recognition system that must recognize both old and new users.
Class Incremental Learning (CIL) is an innovative paradigm designed to enable models to continually expand their knowledge to include new classes without forgetting the old ones. While CIL has made significant strides in the realm of computer vision (CV), its application in audio signal processing has been lagging. Based on these observations, we believe that introducing incremental learning to the field of audio signal processing, i.e., Audio Class-Incremental Learning (AuCIL), is a meaningful endeavor.


To align audio signal processing algorithms with real-world scenarios and strengthen research in audio class-incremental learning, it is crucial to offer a straightforward and efficient toolbox that includes mainstream CIL methods. With this goal in mind and inspired by \cite{zhou2023pycil}, we have developed such a toolbox \textbf{AudioCIL} using the Python programming language, which is widely adopted within the research community. To the best of our knowledge, this is the first incremental learning algorithms library for audio signal processing.

The \textbf{AudioCIL} algorithm library leverages the power of Python to make CIL accessible to the audio signal processing community. Drawing inspiration from CIL advancements in CV, it includes implementations of classic CIL works from the field of CV and offers state-of-the-art algorithms for AuCIL, enabling novel fundamental research.
For user convenience, the core components of \textbf{AudioCIL}, such as data management, model construction, model training, and incremental learning task setting, are modularized within our library. This design allows users to flexibly combine different modules to suit their specific needs.
Furthermore, our toolbox relies solely on standard open-source libraries, making it compatible with a wide range of operating systems, including Linux, macOS, and Windows. The source code for \textbf{AudioCIL} is available on GitHub at \hyperlink{https://github.com/colaudiolab/AudioCIL}{https://github.com/colaudiolab/AudioCIL}

\begin{definition}
(Audio Class-Incremental Learning, AuCIL). AuCIL focuses on learning incrementally from a continuous stream of data that encompasses different classes, while maintaining an effective balance between retaining old knowledge and acquiring new information. We assume there is a total of $T$ training tasks, with each task introducing non-overlapping classes.
Mathematically, we denote the training data for each task as $\mathcal{D}_0, \mathcal{D}_1, \ldots, \mathcal{D}_{T-1}$, where $\mathcal{D}_i \cap \mathcal{D}_j = \varnothing$ for $i \neq j$. For the $i$-th task's training data $\mathcal{D}i$, we represent it as $(X_i, Y_i) = {(x_j, y_j)}_{j=1}^{N_i}$. Here, $x_j \in \mathbb{R}^D$ is a training sample belonging to class $y_j \in Y_i$, and $Y_i$ represents the label space for task $i$. During the training of task $i$, access to data is limited to $\mathcal{D}_i$ only. After each task, the model is evaluated on all classes encountered so far, denoted as $\mathcal{Y}_i = Y_1 \cup Y_2 \cup \ldots \cup Y_i$.
\end{definition}

\begin{definition}
        (Replay buffer). In the $i$-th incremental task, some CIL methods update models only with the current dataset $\mathcal{D}_i$, leading to catastrophic forgetting. To address this, some CIL methods incorporate a replay buffer to store representative samples from past classes. These samples are then integrated with $\mathcal{D}_i$ during model updates, enabling the model to refine its parameters while retaining knowledge of earlier classes.
\end{definition}

\begin{definition}
(Few-shot Audio Class-Incremental Learning, FAuCIL). FAuCIL presents a formidable challenge within the domain of audio signal processing, requiring models to learn incrementally from limited data. Unlike traditional AuCIL, FAuCIL is characterized by an $N$-way-$K$-shot learning paradigm, where each new incremental task introduces $N$ new classes, each represented by a mere $K$ training samples. Typically, $K$ is quite small, often fewer than 20, which underscores the difficulty of the learning task.
\end{definition} 

\begin{figure*}[!t]
\centerline{\includegraphics[width=\linewidth]{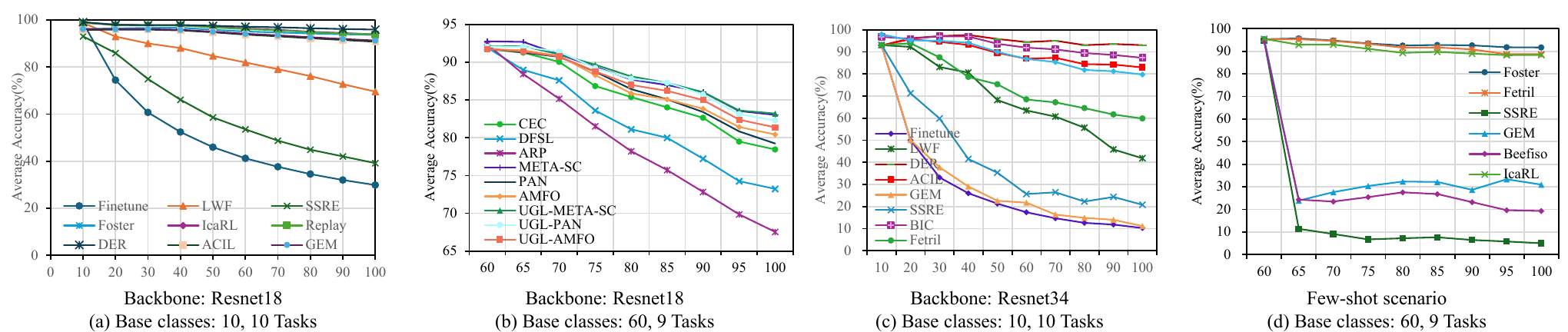}}
\caption{Reproduced incremental accuracy on LS-100 under different settings.}
\label{results}
\end{figure*}

\emph{Implemented Algorithms}. In \textbf{AudioCIL}, we have implemented a total of 16 classic CIL algorithms in CV and 3 state-of-the-art AuCIL algorithms. Here's a succinct overview of each: 
\textbf{Finetune:} Updates model with new task data, prone to catastrophic forgetting.
\textbf{Replay:} Updates model with a mix of new data and samples from a replay buffer.
\textbf{EWC~\cite{kirkpatrick2017overcoming}:} Uses Fisher Information Matrix for regularization against forgetting.
\textbf{LwF~\cite{li2017learning}:}Aligns old and new model outputs via knowledge distillation.
\textbf{iCaRL~\cite{rebuffi2017icarl}:} Retains exemplars from past classes and uses a nearest center mean classifier.
\textbf{GEM~\cite{lopez2017gradient}:} Includes exemplars in gradient updating.
\textbf{BiC~\cite{wu2019large}:} Adds an adaptation layer to iCaRL for new class logits adjustment.
\textbf{WA~\cite{zhao2020maintaining}:} Normalizes weights post-learning, following iCaRL principles.
\textbf{POD-Net~\cite{douillard2020podnet}:} Constrains network with pooled outputs distillation.
\textbf{DER~\cite{yan2021dynamically}:} Features a two-stage, dynamically expandable representation for incremental learning.
\textbf{Coil~\cite{zhou2021co}:} Facilitates bidirectional knowledge transfer using optimal transport.
\textbf{ACIL~\cite{zhuang2022acil}:} Replaces neural network classifier with an analytic one, preserving prior knowledge.
\textbf{META-SC~\cite{li2023fewSC}:} Combines a fixed feature extractor for general representation with a stochastic classifier for new classes.
\textbf{PAN~\cite{li2023few}:} Decouples feature extraction from class prototype adaptation and classifier expansion.
\textbf{AMFO~\cite{li2024few}:} Separates features into task-general and task-specific layers, with a fusion module for balance.

\textbf{Dependencies:} \textbf{AudioCIL} leverages open-source libraries like librosa/torchaudio for audio spectrogram extraction, NumPy and SciPy for linear algebra and optimization tasks, and PyTorch for designing the network architecture.

\textbf{Basic Usage:} \textbf{AudioCIL} offers implementations for 19 class-incremental learning methods. For benchmarking, it supports the LS-100 and NSynth-100 datasets. Users can customize \textbf{AudioCIL} by adjusting global parameters and algorithm-specific hyperparameters before executing the main function. Key global parameters include: 
\textbf{Memory-Size:} Specifies the capacity of the replay buffer used in the incremental learning process. 
\textbf{Init-Cls:} Determines the number of classes in the initial incremental stage. 
\textbf{Increment:} The number of classes in each incremental stage $i, i \geq 1$. 
\textbf{Convnet-type:} Selects the backbone network for the incremental model.
\textbf{Seed:} Establishes the random seed for shuffling class orders, with a default value of 1993.
\textbf{IsFew-shot:} Specifies if the task scenario involves a few-shot learning setting.
\textbf{Kshot:} Defines the number of samples per category in the few-shot learning scenario.

\emph{Evaluation:} The standard metric for evaluating AuCIL is the test accuracy after each task, denoted as $A_i$ , where $i$ represents the task index. Additionally, the average accuracy across all tasks is a widely used measure, calculated as $\mathcal{\hat{A}}=\frac{1}{B}\Sigma_{i=1}^{T}\mathcal{\hat{A}}_i$. 
In our preliminary research within the AuCIL domain, we assessed the incremental performance in terms of Top-1 accuracy throughout the tasks, as depicted in Figure~\ref{results}\footnote{Given the wide variety of algorithms and the performance differences among them, some results tend to overlap when presented in the same chart. Therefore, we selected representative results for each setting and combined the algorithms from different settings to create the algorithm library we have implemented.}. Utilizing benchmark datasets LS-100 and NSynth-100, we divided the 100 classes into multiple incremental tasks. Given that some parameters were not specified in the original paper, we optimized our parameter set in the re-implementation process, including explore the impact of backbone, different task settings. The majority of the reproduced algorithms achieved performance on par with or superior to the results reported in the original paper.

\emph{Conclusion:} We introduce \textbf{AudioCIL}, a Python-based toolbox designed for class-incremental learning in audio classification. It encompasses both foundational and cutting-edge algorithms, facilitating novel research in the field. The consistency in code structure renders \textbf{AudioCIL} a user-friendly tool suitable for research, educational purposes, and industrial applications.



\bibliographystyle{ieeetr}
\bibliography{ref}

\begin{thebibliography}{10}

\bibitem{zhou2023pycil}
D.-W. Zhou, F.-Y. Wang, H.-J. Ye, and D.-C. Zhan, ``Pycil: a python toolbox for
  class-incremental learning,'' {\em SCIENCE CHINA Information Sciences},
  vol.~66, no.~9, p.~197101, 2023.

\bibitem{kirkpatrick2017overcoming}
J.~Kirkpatrick, R.~Pascanu, N.~Rabinowitz, J.~Veness, G.~Desjardins, A.~A.
  Rusu, K.~Milan, J.~Quan, T.~Ramalho, A.~Grabska-Barwinska, {\em et~al.},
  ``Overcoming catastrophic forgetting in neural networks,'' {\em Proceedings
  of the national academy of sciences}, vol.~114, no.~13, pp.~3521--3526, 2017.

\bibitem{li2017learning}
Z.~Li and D.~Hoiem, ``Learning without forgetting,'' {\em IEEE transactions on
  pattern analysis and machine intelligence}, vol.~40, no.~12, pp.~2935--2947,
  2017.

\bibitem{rebuffi2017icarl}
S.-A. Rebuffi, A.~Kolesnikov, G.~Sperl, and C.~H. Lampert, ``icarl: Incremental
  classifier and representation learning,'' in {\em Proceedings of the IEEE
  conference on Computer Vision and Pattern Recognition}, pp.~2001--2010, 2017.

\bibitem{lopez2017gradient}
D.~Lopez-Paz and M.~Ranzato, ``Gradient episodic memory for continual
  learning,'' {\em Advances in neural information processing systems}, vol.~30,
  2017.

\bibitem{wu2019large}
Y.~Wu, Y.~Chen, L.~Wang, Y.~Ye, Z.~Liu, Y.~Guo, and Y.~Fu, ``Large scale
  incremental learning,'' in {\em Proceedings of the IEEE/CVF conference on
  computer vision and pattern recognition}, pp.~374--382, 2019.

\bibitem{zhao2020maintaining}
B.~Zhao, X.~Xiao, G.~Gan, B.~Zhang, and S.-T. Xia, ``Maintaining discrimination
  and fairness in class incremental learning,'' in {\em Proceedings of the
  IEEE/CVF conference on computer vision and pattern recognition},
  pp.~13208--13217, 2020.

\bibitem{douillard2020podnet}
A.~Douillard, M.~Cord, C.~Ollion, T.~Robert, and E.~Valle, ``Podnet: Pooled
  outputs distillation for small-tasks incremental learning,'' in {\em Computer
  vision--ECCV 2020: 16th European conference, Glasgow, UK, August 23--28,
  2020, proceedings, part XX 16}, pp.~86--102, Springer, 2020.

\bibitem{yan2021dynamically}
S.~Yan, J.~Xie, and X.~He, ``Der: Dynamically expandable representation for
  class incremental learning,'' in {\em Proceedings of the IEEE/CVF conference
  on computer vision and pattern recognition}, pp.~3014--3023, 2021.

\bibitem{zhou2021co}
D.-W. Zhou, H.-J. Ye, and D.-C. Zhan, ``Co-transport for class-incremental
  learning,'' in {\em Proceedings of the 29th ACM International Conference on
  Multimedia}, pp.~1645--1654, 2021.

\bibitem{zhuang2022acil}
H.~Zhuang, Z.~Weng, H.~Wei, R.~Xie, K.-A. Toh, and Z.~Lin, ``Acil: Analytic
  class-incremental learning with absolute memorization and privacy
  protection,'' {\em Advances in Neural Information Processing Systems},
  vol.~35, pp.~11602--11614, 2022.

\bibitem{li2023fewSC}
Y.~Li, W.~Cao, J.~Li, W.~Xie, and Q.~He, ``Few-shot class-incremental audio
  classification using stochastic classifier,'' {\em arXiv preprint
  arXiv:2306.02053}, 2023.

\bibitem{li2023few}
Y.~Li, W.~Cao, W.~Xie, J.~Li, and E.~Benetos, ``Few-shot class-incremental
  audio classification using dynamically expanded classifier with
  self-attention modified prototypes,'' {\em IEEE Transactions on Multimedia},
  vol.~26, pp.~1346--1360, 2023.

\bibitem{li2024few}
Y.~Li, J.~Li, Y.~Si, J.~Tan, and Q.~He, ``Few-shot class-incremental audio
  classification with adaptive mitigation of forgetting and overfitting,'' {\em
  IEEE/ACM Transactions on Audio, Speech, and Language Processing}, 2024.

\end{thebibliography}
\end{multicols}
\end{document}